# Wave Packet Dynamics in the Majorana Equation

*Luca Nanni

*corresponding author e-mail: luca.nanni@student.unife.com

### Abstract

In the Majorana equation for particles with arbitrary spin, wave packets occur due to not only the uncertainty that affects position and momentum but also due to infinite components with decreasing mass that form the Majorana spinor. In this paper, we prove that such components contribute to increase the spreading of wave packets. Moreover, Zitterbewegung takes place in both the time propagation of Dirac wave packets and in Majorana wave packets. However, it shows a peculiar *fine structure*. Finally, group velocity always remains subluminal and contributions due to infinite components decrease progressively as spin increases.

## 1  Introduction

The study of the relativistic wave packets of free particles is one of the most investigated topics in quantum mechanics, both for pedagogical reasons [1-3] and for its implication in some theoretical research fields [4-5]. In fact, relativistic wave packets show particularities and peculiarities that do not occur for non-relativistic ones obtained from the Schrödinger equation. The latter spread during time evolution and maintain their Gaussian profile [6], while relativistic wave packets spread and lose their characteristic initial bell-shaped distribution [2,6]. Moreover, the negative energy solutions of the Dirac equation lead to Zitterbewegung [7], a hypothetical wiggling motion of the position of an elementary particle around its mean value. Schrödinger studied this motion in 1930 and interpreted it as a rapid motion that occurs due to the overlapping of the positive and negative energy solutions of the Dirac equation in position space [8]. More recent research has confirmed that this motion arises not necessarily due to the interference between positive and negative energy solutions, but more likely due to the Dirac wave function's complex phase factor [9]. This opens the way to introduce Zitterbewegung also in the solution of quantum equations that have only one energy sign.

This study seeks to investigate the propagation of a fermionic free particle's wave packet in the Majorana equation with infinite components [10]. The study proves that wave packets occur due to the contribution of all infinite components with decreasing mass spectrums, which correspond to increasing spin. Each of these components propagates over time just as a single Dirac wave packet does. Moreover, although the bradyonic solutions of the Majorana equation have only positive energies, Zitterbewegung still occurs. This may transpire solely due to the opposite sign of their complex phase factors. Because of the contribution of all the spinor's infinite components, Majorana wave packet spreading increases in comparison with that of Dirac. The same infinite components lead to an anomalous *fine structure* of Zitterbewegung. In addition, group velocity always remains subluminal, as demonstrated below. Particularly, group velocities associated with solutions involving increasing spin progressively decrease in proportion to the mass spectrum of the bradyonic tower.

## 2   Wave Packets in the Dirac Equation

Many papers address the study of relativistic wave packets in the Dirac equation [1-3, 11-13]. In this section, we retrace in detail the work of Theller [6]. This will serve as a methodological approach for investigating relativistic Majorana wave packets [10]. In the Dirac equation, the Fourier integral of the overlapping between the positive and negative energy solutions produces wave packets in one spatial dimension:

$$\Psi(x,t) = \int_{-\infty}^{\infty}[a(p)\varphi_+(p,x,t) + b(p)\varphi_-(p,x,t)]dp \tag{1}$$

The coefficients $a(p)$ and $b(p)$ are functions of the impulse, while $\varphi_+$ and $\varphi_-$ are the solutions with positive and negative energy:

$$\begin{cases} \varphi_+(p,x,t) = \frac{1}{\sqrt{2\pi}} u_+(p) exp\left\{-i\frac{(px-Et)}{\hbar}\right\} \\ \varphi_-(p,x,t) = \frac{1}{\sqrt{2\pi}} u_-(p) exp\left\{i\frac{(px-Et)}{\hbar}\right\} \end{cases} \tag{2}$$

In **(2)**, $u_+$ and $u_-$ refer to the Dirac spinors of the particle and antiparticle. Note that for one-dimensional motion there is not spin-flip (up-down inversion of the spin). Particles and antiparticles have the same spin component (either up or down), and then $u_+$ and $u_-$ are two component vectors. The functions $a(p)$ and $b(p)$ are the Fourier transforms of the wave packet at the time $t = 0$:

$$\begin{cases} a(p) = \frac{1}{\sqrt{2\pi\hbar}} \int_{-\infty}^{\infty} \Psi(x,0) e^{-ipx/\hbar} dx \left[\frac{1}{2}\left(\mathbb{1}m_0 c^2 - \frac{\mathbb{H}_0(p)}{E_+}\right)\right] \\ b(p) = \frac{1}{\sqrt{2\pi\hbar}} \int_{-\infty}^{\infty} \Psi(x,0) e^{-ipx/\hbar} dx \left[\frac{1}{2}\left(\mathbb{1}m_0 c^2 - \frac{\mathbb{H}_0(p)}{E_-}\right)\right] \end{cases} \tag{3}$$

In **(3)**, $\mathbb{1}$ is the 2x2 unitary matrix and $\mathbb{H}_0(p)$ is the 2x2 Hamiltonian matrix:

$$\mathbb{H}_0(p) = \begin{pmatrix} m_0 c^2 & cp \\ cp & -m_0 c^2 \end{pmatrix} \tag{4}$$

Also, $E_\pm$ signifies the energies of the positive and negative solutions that the relativistic relation $\pm\sqrt{p^2c^2 + m_0^2c^4}$ presents. In this study, we hypothesize that the initial plan wave is of a Gaussian type:

$$\Psi(x,0)_\pm = \frac{1}{\sqrt{2\pi\sigma^2}} \exp\left\{-\frac{x^2}{2\sigma^2}\right\} u_\pm \qquad (5)$$

At $t = 0$, the particle position is within the range $\pm\sigma$ with a zero mean value. Substituting equation **(5)** in **(3)** and performing the Fourier integral produces:

$$\begin{cases} a(p) = \frac{\sigma}{2\sqrt{\pi\hbar}} \left(\mathbb{1}m_0c^2 - \frac{\mathbb{H}_0(p)}{E_+}\right) e^{-\sigma^2 p^2} \\ b(p) = \frac{\sigma}{2\sqrt{\pi\hbar}} \left(\mathbb{1}m_0c^2 - \frac{\mathbb{H}_0(p)}{E_-}\right) e^{-\sigma^2 p^2} \end{cases} \qquad (6)$$

Let us consider the two matrices in brackets; an easy calculation yields their explicit form:

$$\left(\mathbb{1}m_0c^2 - \frac{\mathbb{H}_0(p)}{E_\pm}\right) = \begin{pmatrix} 1 \mp \frac{1}{\gamma} & \mp 1 \\ \mp 1 & 1 \pm \frac{1}{\gamma} \end{pmatrix} \qquad (7)$$

In **(7)**, $\gamma$ is the Lorentz relativistic factor. Since we are interested in the dynamics of the particle close to the speed of light, we can simplify **(7)** as:

$$\left(\mathbb{1}m_0c^2 - \frac{\mathbb{H}_0(p)}{E_\pm}\right)_{v\to c} = \begin{pmatrix} 1 & \mp 1 \\ \mp 1 & 1 \end{pmatrix} \qquad (8)$$

Replacing **(2)**, **(6)** and **(8)** in **(1)** and considering the column vectors $u_+ = (1,0)^t$, $u_- = (0,1)^t$ produces the wave packet spinor:

$$\Psi(x,t) = \frac{\sigma}{2\sqrt{2\pi^2\hbar}} \left[ \begin{pmatrix} 1 \\ -1 \end{pmatrix} e^{-iEt/\hbar} \int_{-\infty}^{\infty} e^{-\sigma^2 p^2} e^{-ipx/\hbar} dp + \begin{pmatrix} 1 \\ 1 \end{pmatrix} e^{iEt/\hbar} \int_{-\infty}^{\infty} e^{-\sigma^2 p^2} e^{ipx/\hbar} dp \right] \qquad (9)$$

The first and last integrals represent respectively the Fourier transform and anti-transform of the function $e^{-\sigma^2 p^2}$:

$$T(e^{-\sigma^2 p^2}) = \frac{1}{\sqrt{8\sigma}} e^{-x^2/\sigma^2}, \quad T^{-1}(e^{-\sigma^2 p^2}) = \frac{1}{\sqrt{2\pi\sigma^2}} e^{-x^2/2\sigma^2} \qquad (10)$$

Substituting **(10)** in **(9)** presents the final form of the wave packet in one-dimensional space, the components of which are:

$$\Psi(x,t) = \begin{cases} \frac{1}{8\pi\sqrt{\hbar}} e^{-iEt/\hbar} e^{-(x^2/\sigma^2)} + \frac{1}{4\pi\sqrt{\pi\hbar}} e^{iEt/\hbar} e^{-(x^2/2\sigma^2)} \\ -\frac{1}{8\pi\sqrt{\hbar}} e^{-iEt/\hbar} e^{-(x^2/\sigma^2)} + \frac{1}{4\pi\sqrt{\pi\hbar}} e^{iEt/\hbar} e^{-(x^2/2\sigma^2)} \end{cases} \qquad (11)$$

The spinor **(11)** shows that the two components are linear combinations of Gaussians with different widths, modulated by a time-oscillating function. As soon as $t \neq 0$, the Gaussians $e^{-(x^2/\sigma^2)}$ and $e^{-(x^2/2\sigma^2)}$ begin translating in opposite directions by an amount equal to $Et/\hbar$ and lose their symmetrical shapes. If we replace the explicit form of relativistic energy in the term $Et/\hbar$ and consider that $x = ct$ we get the translation $T$:

$$T = \pm\gamma \frac{m_0 c}{\hbar} \qquad (12)$$

In **(12)**, the positive and negative signs refer respectively to the first and second components of the wave packet **(11)**. Depending on the Lorentz factor, the translation **(12)** increases with the velocity of the particle.

As mentioned in the introduction [7-8], the interference of the two positive and negative energy solutions leads to Zitterbewegung. This motion occurs at the speed of light and continues as long as the positive and negative energy solutions overlap in position space. Therefore, the ripples characterizing the motion of the particle's position around its mean value fade away as soon as the particle begins to evolve over time.

During time evolution, the wave packet spreads more and more in accordance with the functional relationship between the variance and the propagation time [14]:

$$\sigma = \sigma_0 \sqrt{1 + \frac{2\hbar^2 t^2}{m_0^2 \sigma_0^2}} \qquad (13)$$

Finally, the next equation shows the group velocity of the relativistic Dirac wave packet:

$$v_g = \frac{\partial E}{\partial p} = \frac{\partial}{\partial p} \sqrt{p^2 c^2 + m_0^2 c^4} = \frac{c}{\sqrt{2-\beta^2}} \leq c \qquad (14)$$

In **(14)**, $\beta = v/c$. The group velocity always remains subluminal and tends towards $c$ as the particle velocity approaches the speed of light.

## 3   Wave Packets in the Majorana Equation

Let us now study wave packets in the Majorana equation using the same approach adopted in the previous section. Consider a free particle with half-integer spin $s_0$ and rest mass $m_0$. The Majorana equation returns a solution with infinite components given by the linear combination of the ground state and all of the infinite excited states with increasing intrinsic angular momentum [10,15]:

$$\varphi_\pm(n,p,x,t) = \frac{1}{\sqrt{2\pi}} u_\pm(n,p) \exp\left\{\mp i \frac{(p_n x - E_n t)}{\hbar}\right\} \qquad (15)$$

In **(15)**, $n$ is the positive integer corresponding to the excited state in which the following formula presents the mass and spin [16]:

$$\begin{cases} m(n) = m_0 / \left(\frac{1}{2} + J_n\right) \\ J_n = s_0 + n \end{cases} \qquad (16)$$

This, in turn, shows the momentum and energy as [16]:

$$\begin{cases} p(n) = \gamma \frac{m_0}{(n+1)} v \\ E(n) = \gamma \frac{m_0}{(n+1)} c^2 \end{cases} \qquad (17)$$

The occupation probability of the nth excited state is [16]:

$$P_n = \sqrt{(v/c)^n - (v/c)^{n+1}} \qquad (18)$$

Equations **(17)** and **(18)** establish that when a particle's velocity approaches the speed of light, the excited states with high spin $J_n$ become stable. Another peculiarity of the Majorana equation is that all bradyonic solutions have positive energies [10]. Therefore, in view of the Majorana theory of particles with arbitrary spin, the wave function **(1)** becomes:

$$\Psi(x,t) = \int_{-\infty}^{\infty} \sum_{n=0}^{\infty} P_n [a(n,p)\varphi_+(n,p,x,t) + b(n,p)\varphi_-(n,p,x,t)] dp \tag{19}$$

In **(19)**, every excited state is weighted for its occupation probability. Since all solutions (for both particles and antiparticles) have positive energies, the functional coefficients $a(n,p)$ and $b(n,p)$ are equal:

$$a(n,p) = b(n,p) = \frac{1}{\sqrt{2\pi\hbar}} \int_{-\infty}^{\infty} \Psi(x,0) e^{-ip_n x/\hbar} dx \left[ J_n \left( \mathbb{1} m_0 c^2 - \frac{\mathbb{H}_n(p)}{E(n)} \right) \right] \tag{20}$$

Proceeding like Dirac, the matrix $\left( \mathbb{1} m_0 c^2 - \frac{\mathbb{H}_n(p)}{E(n)} \right)$ becomes:

$$\left( \mathbb{1} m_0 c^2 - \frac{\mathbb{H}_n(p)}{E(n)} \right) = \begin{pmatrix} 1 - \frac{n+1}{\gamma} & -1 \\ -1 & 1 - \frac{n+1}{\gamma} \end{pmatrix} \tag{21}$$

Unlike the cases in the previous section, this case cannot simplify **(21)**, because as the particle approaches the speed of light, both $\gamma$ and $n$ increase. Substituting **(20)** and **(21)** in **(19)** and using the spin vectors $u_+ = (1,0)^t$, $u_- = (0,1)^t$ gives us:

$$\Psi(x,t) = \begin{cases} \sum_{n=0}^{\infty} P_n J_n \left[ \frac{1-(n+1)/\gamma}{8\pi\sqrt{\hbar}} e^{-iE_n t/\hbar} e^{-(x^2/\sigma^2)} - \frac{1}{4\pi\sqrt{\pi\hbar}} e^{iE_n t/\hbar} e^{-(x^2/2\sigma^2)} \right] \\ \sum_{n=0}^{\infty} P_n J_n \left[ \frac{1}{4\pi\sqrt{\pi\hbar}} e^{iE_n t/\hbar} e^{-(x^2/\sigma^2)} - \frac{1-(n+1)/\gamma}{8\pi\sqrt{\hbar}} e^{-iE_n t/\hbar} e^{-(x^2/2\sigma^2)} \right] \end{cases} \tag{22}$$

For $J_n = 1/2$ and for slow motion, only the ground state is occupied, and **(22)** becomes:

$$\Psi(x,t) = \begin{cases} \frac{1}{4\pi\sqrt{\pi\hbar}} e^{iE_0 t/\hbar} e^{-(x^2/2\sigma^2)} \\ -\frac{1}{4\pi\sqrt{\pi\hbar}} e^{iE_0 t/\hbar} e^{-(x^2/2\sigma^2)} \end{cases} \tag{23}$$

In other words, the two components of Majorana wave packets are opposite Gaussian functions translating with the same velocity $m_0 c/\hbar$. When $v \gg 0$ then $J_n > 1/2$ and $n > 1$, and each component of the spinor **(22)** is a linear combination of infinite Gaussian functions weighted for the coefficient $P_n$. In particular, the first component is always negative while the latter is always positive. In spacetime, each Gaussian of the linear combination **(22)** translates to a quantity that the following provides:

$$T_n = \pm \gamma \frac{m_0 c}{\hbar(n+1)} \tag{24}$$

The Majorana wave packet fades away in all of its infinite components, spreading faster than that of Dirac. Note that each component of the spinor propagates with its own velocity. In fact, using **(16)** yields:

$$v_T(n) = \frac{m_0 c}{\left(\frac{1}{2} + s_0 + n\right)} \tag{25}$$

This shows that the propagation velocity decreases with order $n$ of the excited state. Since the Majorana solutions have only positive energies, Zitterbewegung occurs due to the different sign of the complex phase factor. Equation **(25)** proves that all infinite components translate with different velocities. Thus, Zitterbewegung shows that small ripples form a *fine structure* due to the progressive decreases of the overlapping of infinite Gaussian functions. This behaviour does not appear in Dirac wave packets. The occupation probability of excited states increases with particle velocity. Therefore, the slower the particle motion, the faster the *fine structure* fades away. If Zitterbewegung is a peculiarity of the relativistic dynamics of the wave packet, its *fine structure* increases even further.

In the Majorana picture, the relationship between variance and propagation time is very similar to **(13)**. However, an additional equation represents states with increasing spin:

$$\sigma = \sigma_0 \sqrt{1 + \frac{2(n+1)\hbar^2 t^2}{m_0^2 \sigma_0^2}} \tag{26}$$

From **(26)**, we conclude that wave packet components related to excited states spread faster as their order $n$ rises. This confirms the comment above about **(24)**.

Let us now calculate the group velocity of Majorana wave packets. To do this, we have to consider the fact that it will depend on the occupation probability of the nth excited state. For that reason, we can write the group velocity as:

$$v_g^M(n) = v_g \sqrt{(v/c)^n - (v/c)^{n+1}} = v_g \sqrt{1 - (v/c)}(v/c)^{n/2} \tag{27}$$

In **(27)**, $v_g$ refers to **(14)** and superscript *M* refers to Majorana theory. Replacing the explicit form of **(14)** in **(27)** yields:

$$v_g^M(n) = \frac{c}{\sqrt{2-\beta^2}} \sqrt{1-\beta}(\beta)^{n/2} \tag{28}$$

The term $\sqrt{1-\beta}(\beta)^{n/2}$ is always lower than one and tends towards zero as the particle's velocity approaches the speed of light. That means states with high spin have a group velocity tending towards zero. This is further confirmation that Majorana wave packets spread faster than Dirac wave packets do. The relativistic composition of all single terms **(28)** yields the overall group velocity of Majorana wave packets, which always remains subluminal. We can easily perform this composition in the reference frame of the centre of mass for the particle, where the occupied state is the fundamental one [10]. Equation **(28)** also shows that for slow motion, both $\beta$ and $n$ tend towards zero, and group velocity becomes equal to that of a Dirac wave packet.

# 4   Conclusion

This study shows that when we account for free particles with arbitrary spin in the picture of Majorana theory, relativistic wave packets spread over time. This occurs not only because of the uncertainty that affects position and momentum but also due to the infinite components of spinors with decreasing masses. During propagation, these components translate with different velocities. Their overlapping in position space gradually decreases to zero, which leads to the formation of a *fine structure* in Zitterbewegung. Since occupation probabilities of spinor components decrease when their order $n$ increases, the *fine structure* fades away very quickly. This is considerable only when the particle velocity approaches the speed of light. This result is completely unexpected and, even if not physically observable, represents an additional breaking point between the Dirac and Majorana theories. We also demonstrated that the contribution of single group velocities of infinite components yields the overall group velocity. This value decreases when their order $n$ increases, and always remains subluminal.

The results of this theoretical work could facilitate the comprehension of relativistic motions of fermionic particles with any value of spin, especially in cases where the material wave interacts with a finite potential barrier. This last topic will be the subject of a new study whose purpose is to investigate the behaviour of Majorana fermions in tunnelling phenomena.